\documentclass[12pt]{article}
\usepackage{epsfig}

\newcommand{\be}{\begin{eqnarray}}
\newcommand{\ee}{\end{eqnarray}}
\newcommand{\sll}{\raise.15ex\hbox{$/$}\kern-.43em\hbox{$l$}}
\newcommand{\slp}{\raise.15ex\hbox{$/$}\kern-.43em\hbox{$p$}}
\newcommand{\slq}{\raise.15ex\hbox{$/$}\kern-.43em\hbox{$q$}}
\newcommand{\slk}{\raise.15ex\hbox{$/$}\kern-.43em\hbox{$k$}}
\newcommand{\slepsilon}{\raise.15ex\hbox{$/$}\kern-.53em\hbox{$\epsilon$}}

\begin{document}

\bibliographystyle{unsrt}
\footskip 1.0cm

\thispagestyle{empty}
\begin{flushright}
INT--PUB 05--21
\end{flushright}
\vspace{0.1in}

\begin{center}{\Large \bf {Photon $+$ Hadron Production in High Energy 
Deuteron (Proton)-Nucleus Collisions}}\\

\vspace{1in}
{\large  Jamal Jalilian-Marian}\\

\vspace{.2in}
{\it Institute for Nuclear Theory, University of Washington,
Seattle, WA 98195\\ }

\end{center}

\vspace*{25mm}

\begin{abstract}

\noindent We apply the Color Glass Condensate formalism to photon $+$ 
hadron production cross section in high energy deuteron (proton)-gold 
collisions at RHIC. We investigate the dependence of the production 
cross section on the angle between the produced hadron and photon for 
various rapidities and transverse momenta. It is shown that the angular
correlation between the produced hadron and photon is a sensitive probe
of the saturation dynamics.

\end{abstract}
\newpage

\section{Introduction}

The recent RHIC results on the suppression of the hadron transverse 
momentum spectra in deuteron-gold collisions, as compared to proton-proton
collisions, in the forward rapidity region \cite{rhic} have caused a major
excitement in the RHIC community. While the observed suppression was
qualitatively predicted by the Color Glass Condensate 
(CGC)\footnote{See \cite{iv} for reviews and extensive references.} based approaches 
\cite{kkt1}, and verified by more quantitative analysis \cite{kkt2,aaj},
there have been newer models introduced recently which can also fit the data 
\cite{rudi}. Measurement of different processes, for example prompt 
photon \cite{al} or heavy quark pair production \cite{bgv}, at RHIC can therefore 
help establish the dominance 
of the saturation dynamics in the forward rapidity region at RHIC.

Unlike hadron production, electromagnetic processes such as prompt photon 
or dilepton \cite{fgjjm}  
production do not involve hadronization of the final state and are therefore
a cleaner process in which to investigate the saturation dynamics. Due 
to smallness of the electromagnetic coupling constant, however, these processes
are rare and require high beam luminasity and/or long running times of
the machine. Nevertheless, in order to clarify the underlying physics of particle production in
forward rapidity region at RHIC and eventually in mid or forward rapidity LHC, 
it is essential to measure photons, dileptons, etc. at RHIC. A measurement of
photon production in the forward rapidity region, for example, can distinguish
between recombination models and the Color Glass Condensate physics since
one does not expect recombination effects to be relevant for prompt photon
production while one expects a suppression pattern for prompt photons very similar to hadrons
in deuteron (proton)-nucleus collisions in the Color Glass Condensate formalism. 

Two particle correlations are known to be a sensitive probe of saturation
dynamics. While the two hadron correlation function in mid rapidity is expected
to broaden due to multiple scattering, it is expected to disappear as one
measures two widely separated (in rapidity) hadrons due to the small $x$ quantum
evolution \cite{klm}. This qualitative expectation is verified by preliminary data from RHIC 
\cite{star}. A more quantitative theoretical analysis of the two hadron correlation
in the Color Glass Condensate formalism \cite{jjmyk} is however quite challenging 
due to presence of $3$ and $4$-point functions of Wilson lines and therefore
requires making approximations (for example, large $N_c$ or Gaussian ansatz) and 
assumptions which may affect the outcome of the analysis. Furthermore, since the
measured hadrons are at low to intermediate transverse momenta, non-perturbative
effects may play an important role. 

Therefore, in this brief note, we consider production of a hadron
and a photon as the cleaner and theoretically simpler process in which to
investigate angular correlations and the role of saturation physics. This process
has the further advantage that it is not expected to suffer from possible  recombination 
effects and can therefore help establish/constrain validity of applying the Color 
Glass Condensate formalism to deuteron-nucleus collisions at RHIC. Experimentally,
it is possible to measure this process at RHIC, using the STAR detector, for example,
even though one will most likely need another deuteron-gold run for improved statistics 
and detector coverage.

As shown in \cite{aaj}, one probes the very large $x$ region in the deuteron wave function
and the very small $x$ region in the target wave function (see Fig. $10$ in \cite{aaj}
for the values of $x$ contributing to single hadron production).  In this kinematics, one
can treat the incoming deuteron (proton) as a dilute object, consisting of quarks and gluons 
\cite{adjjm} 
with a possible nuclear modification of the deuteron wave function which should be very small
for the typical $x$ and $Q^2$ involved.  On the other hand, the values of $x$ probed in
the target wave function are very small, $O( 10^{-3} - 10^{-4})$. Therefore, it is essential
to include the high gluon density effects in the target. 

The quark-photon production cross section in quark-nucleus scattering
has been calculated in the Color Glass Condensate formalism in \cite{fgjjm}.
We use this cross section and convolute it  with quark (anti-quark) distributions 
in a deuteron (proton) and quark-hadron fragmentation function to make predictions
for hadron-photon angular correlations at RHIC and LHC. To probe the smallest
$x$ possible in a given process in a hadron-hadron (nucleus-nucleus) collider
environment, one needs to be in the forward rapidity region. RHIC has the further 
advantage of having a very unique forward rapidity capability, with plans for 
detector upgrades. One therefore can expect to be able to measure the hadron-photon
correlation in a wide kinematic region at RHIC. For example, the STAR collaboration 
will be able to measure hadrons and photons at rapidities $y=0$ and $4$ and study
their correlations.

\section{The Scattering Cross Section}

The scattering cross section for production of a massless on-shell quark with momentum
$\vec{l}$ and a real photon with momentum $\vec{k}$ was derived in  \cite{fgjjm}. It is given
by
\be
&&{d\sigma^{q(p)\, A \rightarrow q(l)\,\gamma(k)\, X}
\over d^2b_t\, dk_t^2\, dl_t^2\, dy_{\gamma}\, dy_l\, d\theta} =
{e_q^2\, \alpha_{em} \over \sqrt{2}(2\pi)^2} \, 
{k^-\over  k_t^2 \sqrt{s}} \,
{1 + ({l^-\over p^-})^2 \over
[k^- \, \vec{l_t} - l^- \vec{k_t}]^2}\nonumber \\
&&\delta [x - {l_t \over \sqrt{s}} e^{y_l} - {k_t \over \sqrt{s}} e^{y_{\gamma}} ] \,
\bigg[ 2 l^- k^-\, l_t\, k_t \,\cos \theta + k^- (p^- -k^-)\, l_t^2 + l^- (p^- -l^-)\, k_t^2 \bigg] \nonumber \\
&&\int dr_t \, r_t \, J_0 [r_t | \vec{l_t} + \vec{k_t}|]    \, N (b_t, r_t, x_g) 
\label{eq:cs_gen}
\ee
where the incoming quark has momentum $p$, the photon and outgoing quark rapidities 
are defined via $k^- ={k_t \over \sqrt{2}} e^{y_{\gamma}}$
and $l^- ={l_t \over \sqrt{2}} e^{y_l}$. The angle $\theta$ is the opening angle between the 
final state quark and photon defined as $cos \, \theta \equiv {l_t \cdot k_t \over  l_t k_t}$, with
respect to the produced quark axis. The dipole cross section $N$ satisfies the JIMWLK
equation and has all the multiple scattering and small $x$ evolution effects encoded. It 
is defined as 
\be
N(b_t,r_t,x_g) = {1\over N_c} \, Tr <1 - V^{\dagger} (x_t) V (y_t)>
\label{eq:cs_def}
\ee
with $b_t\equiv (x_t + y_t)/2$ and $r_t\equiv x_t - y_t$. The dipole cross section depends
on Bjorken $x_g$ via the JIMWLK renormalization group equations. In the present case,
it is related to the photon and final state quark rapidities and transverse momenta via
\be
x_g = {1\over \sqrt{s}}[k_t e^{-y_{\gamma}} + l_t e^{-y_l}]\, .
\label{eq:x_g}
\ee

 In order to compute the 
hadron $+$ photon production cross section, we would need to convolute the above
partonic cross section with the quark and anti-quark distributions of a deuteron (proton)
and quarh-hadron fragmentation function. However, before doing that, it is instructive 
to investigate in some detail, the properties of the above cross section. Specifically, 
we would like to investigate the dependence of this cross section on the angle between
the produced quark and photon. In order to do this, we isolate the parts of the cross
section which depend on the angle and ignore the rest of the kinematic factors for the
moment. Therefore we define the angle dependent part of the cross section as
\be
I (\theta ) \equiv { [ 2 l^- k^-\, l_t\, k_t \,\cos \theta + k^- (p^- -k^-)\, l_t^2 + l^- (p^- -l^-)\, k_t^2 ] 
\over 
[k^- \, \vec{l_t} - l^- \vec{k_t}]^2} \int dr_t  r_t  J_0 [r_t | \vec{l_t} +\vec{k_t}|]   N (b_t, r_t, x_g) 
\label{eq:ang}
\ee
Without making any assumptions about the specific form of the dipole cross section $N$, it is clear 
from the factor ${1\over  [k^- \, \vec{l_t} - l^- \vec{k_t}]^2}$ that $I (\theta)$ diverges
when the momenta of produced quark and photon are parallel ($\theta \rightarrow 0$). 
This is the standard collinear divergence present in perturbation theory and is not 
affected by saturation physics. On the other hand, when $\vec{l_t} = - \vec{k_t} $, then
$| \vec{l_t} + \vec{k_t}| \rightarrow 0$ and the integral over the dipole size $r_t$ is divergent. This is
taken care of by demanding color neutrality of the target which would cut the dipole size off at 
$r_t \sim 1 fm$. It should be noted that this is different from the single hadron production
case where, at finite transverse momentum of the produced hadron, the integral over
dipole size $r_t$ is always finite.

To proceed further, we need to know the dipole profile $N (b_t, r_t,x_g)$. It satisfies the
JIMWLK equation and therefore can be obtained by solving the JIMWLK equation with
a suitable boundary condition. However, this is highly nontrivial and in practice, it is much
more convenient to use one of the available parameterizations such as the KKT profile \cite{kkt2}.
In this model, the dipole profile is given by 
 \be 
N (b_t, r_t, x_g) = \left (\exp\left[ - \frac{1}{4} [r_t^2
  Q_s^2(y)]^{\gamma(y,r_t)}\right] -1 \right )
  \label{eq:cs_kkt}
\ee
where the anomalous dimension $\gamma (y, r_t)$ is 
\be
\gamma(y,r_t) = \frac{1}{2}\left(1+
\frac{\xi(y,r_t)}{\xi(y,r_t)+\sqrt{2\xi(y,r_t)}+28\zeta(3)}
   \right)
   \label{eq:ano_kkt}
\ee
and 
\be
\xi (y,r_t) = \frac{\log 1/r_t^2 Q_0^2}{(\lambda/2)(y-y_0)}~.
\ee
The saturation scale is given by $Q_s(y) = Q_0 \exp [\lambda (y-y_0)/2] $  with 
$y  = \ln 1/x_g$ and $y_0 =0.6, \, \lambda = 0.3$.
This is the form of the dipole profile we will use for our numerical analysis.
However, it is instructive to consider the analytic behavior of $I(\theta)$. This is possible in 
two limits; when $\gamma (y,r_t)$ is either $1/2$ or $1$. The former corresponds to the BFKL
and saturation region while the later is the case when one has the standard DGLAP anomalous 
dimension. It turns out that the case $\gamma = 1/2$ is very close to the actual value of the
anomalous dimension when one does the integral numerically. Therefore, we use this value 
in our analytic estimates. The integral over the dipole size $r_t$ can be then done exactly
and gives 
\be
 \int dr_t  r_t  J_0 [r_t | \vec{l_t} +\vec{k_t}|]   N (b_t, r_t, x_g)|_{\gamma =1/2} =
 {16 \over
 Q_s^2 [1 + {16 |\vec{l_t} + \vec{k_t}|^2 \over Q_s^2}]^{{3\over 2}}}
 \label{eq:N_approx}
 \ee
 Strictly speaking, the integral over $r_t$ is divergent  when $\theta = \pi$ if 
 $|l_t|=|k_t|$ and needs to be regularized by confinement scale $r_t \sim 1 fm$. 
 To avoid the numerical complications of regularization, in this work we stay 
 away from $\theta = \pi $ point and keep in mind that the expression in 
 (\ref{eq:N_approx}) is not valid\footnote{Also, due to the choice of $\gamma =1$,
 the high transverse momentum limit of the dipole cross section is not reproduced correctly,
 but this is of minor importance for $\theta \rightarrow \pi$ since in this limit the magnitude
 of the argument of the dipole cross section, $ |\vec{l_t} + \vec{k_t}| \rightarrow 0$, is
  always less than 
 the saturation scale. We emphasize that in the numerical analysis, the exact form
 of $\gamma$ is kept so that the high transverse momentum limit of the dipole cross section
 is correctly reproduced.}  at $\theta=\pi$. Using this 
 expression in $I (\theta)$ gives
  \be
 I (\theta) \rightarrow  
  {
  k_t \, e^{-y_l + y_{\gamma}}\,  [ \vec{l_t} + \vec{k_t}]^2
\over l_t \, 
 [({k^-\over l^-}) \, \vec{l_t} - \vec{k_t}]^2}
 {1 \over Q_s^2 } \, {1 \over
 [1 + {16 |\vec{l_t} + \vec{k_t}|^2 \over Q_s^2}]^{{3 \over 2}}}
 \label{eq:I_lim}
 \ee
 This form of $I (\theta) $ shows the angular dependence of the cross section most 
 clearly and can serve as a guide understanding the exact numerical results shown later.
 Note that the above expression does not depend on whether $k_t$ is larger or smaller
 than the saturation scale $Q_s$ and is therefore vaild in both the saturation and BFKL
 regions to the extent that  $\gamma \simeq 1/2$ remains a good approximation. 
 
\begin{figure}[htp]
\centering
\setlength{\epsfxsize=8cm}
\centerline{\epsffile{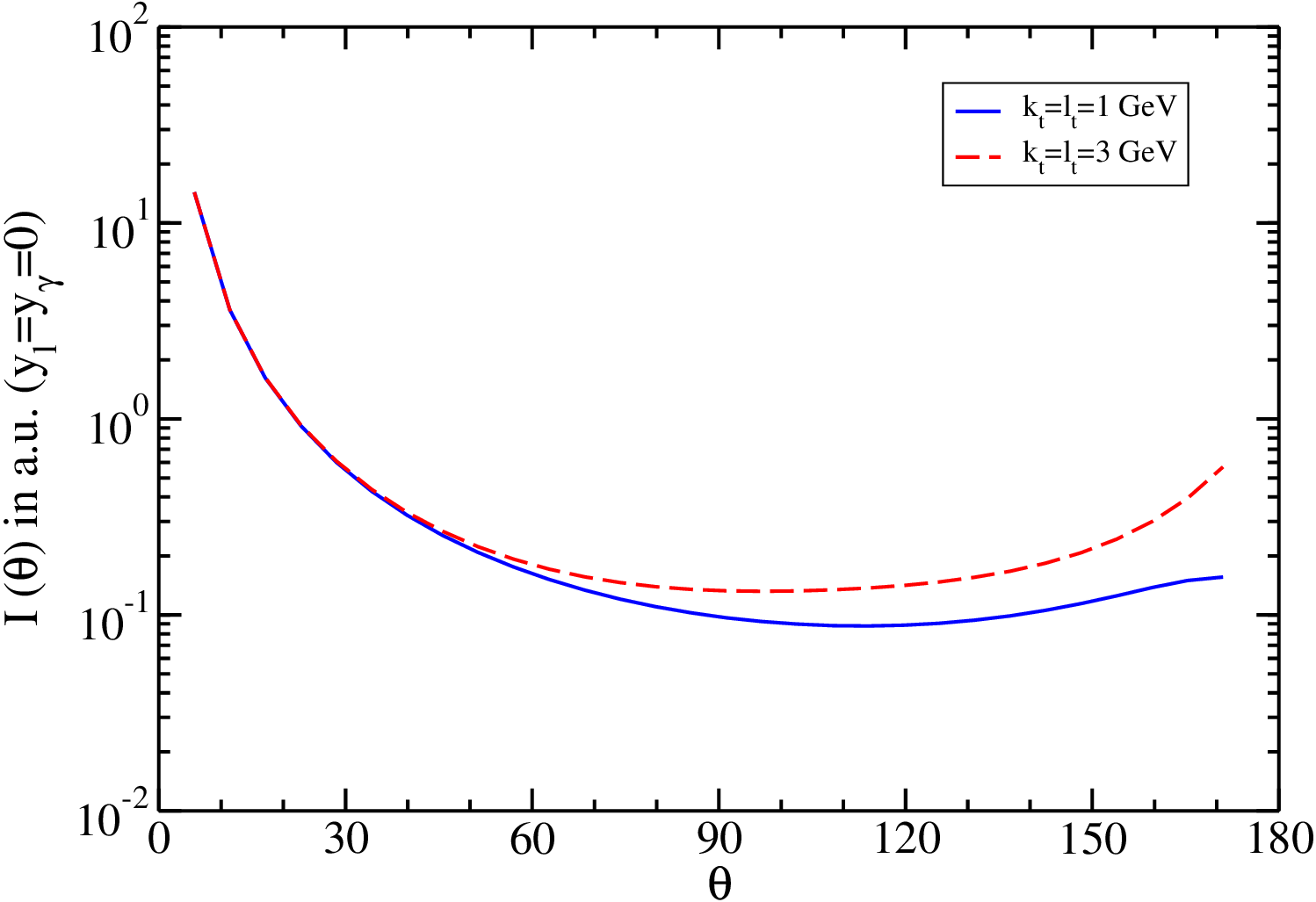}}
\caption{The angular part of the cross section, $I (\theta)$ in mid rapidity RHIC.}
\label{fig:I_theta}
\end{figure}

 In Fig. (\ref{fig:I_theta}) we show the function $I (\theta) $ where both the produced
 photon and quark are in mid rapidity (RHIC) and for the case where both have equal
 transverse momenta. 
% \vspace{0.3in}
 The solid line is when both quark and photon have transverse 
 momenta $l_t = k_t = 1GeV$ and the dashed line is when both transverse momenta
 are $3 GeV$. For ease of comparison, we normalize the two curves at the smallest angle. 
At the smaller momentum $1 GeV$, the "away" side ($\theta \rightarrow 0$) correlation is flat 
 while for higher momentum $3 GeV$, the away side peak is returning. This is mainly
 due to the increase of $x_g$, as given by (\ref{eq:x_g}) and therefore a decrease 
 of $Q_s^2 (x_g)$ which indicates a weakening of the gluon saturation effects in the 
 target nucleus. It should be noted that introduction of an "isolation" procedure 
 experimentally in the photon measurement would cutoff the collinear divergence
 present in $I (\theta)$ which is apparent in the small angle limit in Fig. (\ref{fig:I_theta}).
 Since the isolation cuts needed will depend on the detector geometry after a possible upgrade
 and are not known at the moment, we do not consider an isolation cut on the photon.

\begin{figure}[hbp]
\centering
\setlength{\epsfxsize=8cm}
\centerline{\epsffile{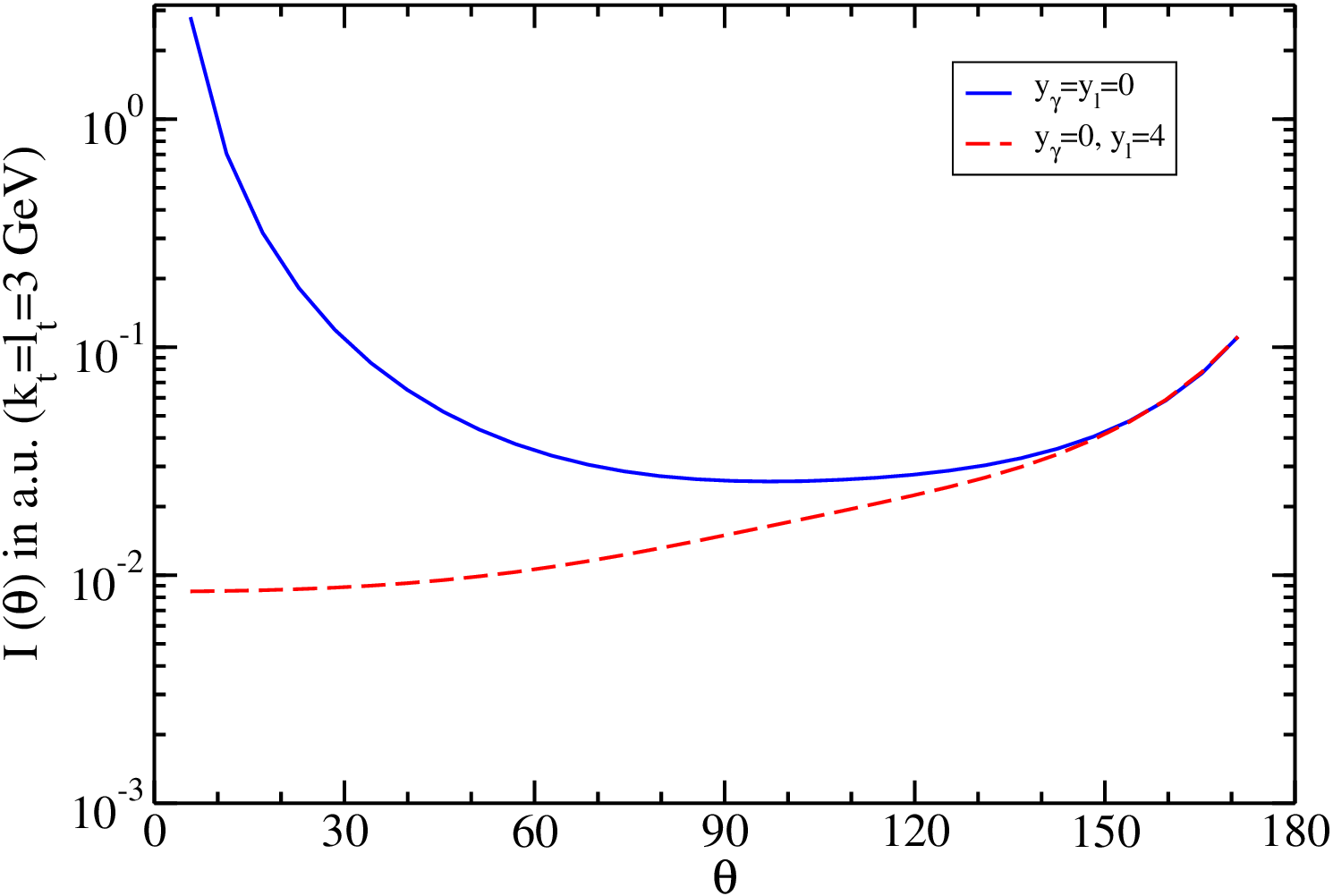}}
\caption{Rapidity dependence of $I (\theta)$ at different transverse momenta.}
\label{fig:I_theta_y}
\end{figure}

In Fig. (\ref{fig:I_theta_y}) we show the angular correlation function $I (\theta)$ at
different rapidities. The solid line shows the case when both particles are in mid 
rapidity while the dashed line is when the produced quark is at $y=4$ while the 
produced photon is in mid rapidity. In both cases, the transverse momenta of both
particles is $3 GeV$. This time, however, the two curves are normalized at the highest
angle. As before, the angular correlation function diverges in mid rapidity when 
$\theta \rightarrow 0$ due to the unregularized collinear divergence.
\vspace{0.3in}
we note that there is no collinear divergence ($\theta \rightarrow 0$) in the case when the rapidity 
separation is so large, $\Delta y =4$, and when $l_t =k_t = 3 GeV$ while the mid rapidity 
correlation function is collinear divergent. 

Finally, we consider the case when both produced partons are in the forward
rapidity region, $y_l = y_{\gamma} = 4$. In this kinematics, one is restricted
to small momenta due to the available phase space so we consider $l_t = k_t =1 GeV$
as well as $l_t = k_t =1.7 GeV$ which is very close to the upper limit allowed by kinematics.
Note that the two curves are not normalized like the previous figures.
Again in the small angle limit, one is probing the collinear divergence present in the 
angular function which shows up as sharp rise of $I (\theta)$. In a realistic experimental
set up, this will be tamed by the isolation cut imposed on the photon. 
\vspace{0.3in}
\begin{figure}[htp]
\centering
\setlength{\epsfxsize=8cm}
\centerline{\epsffile{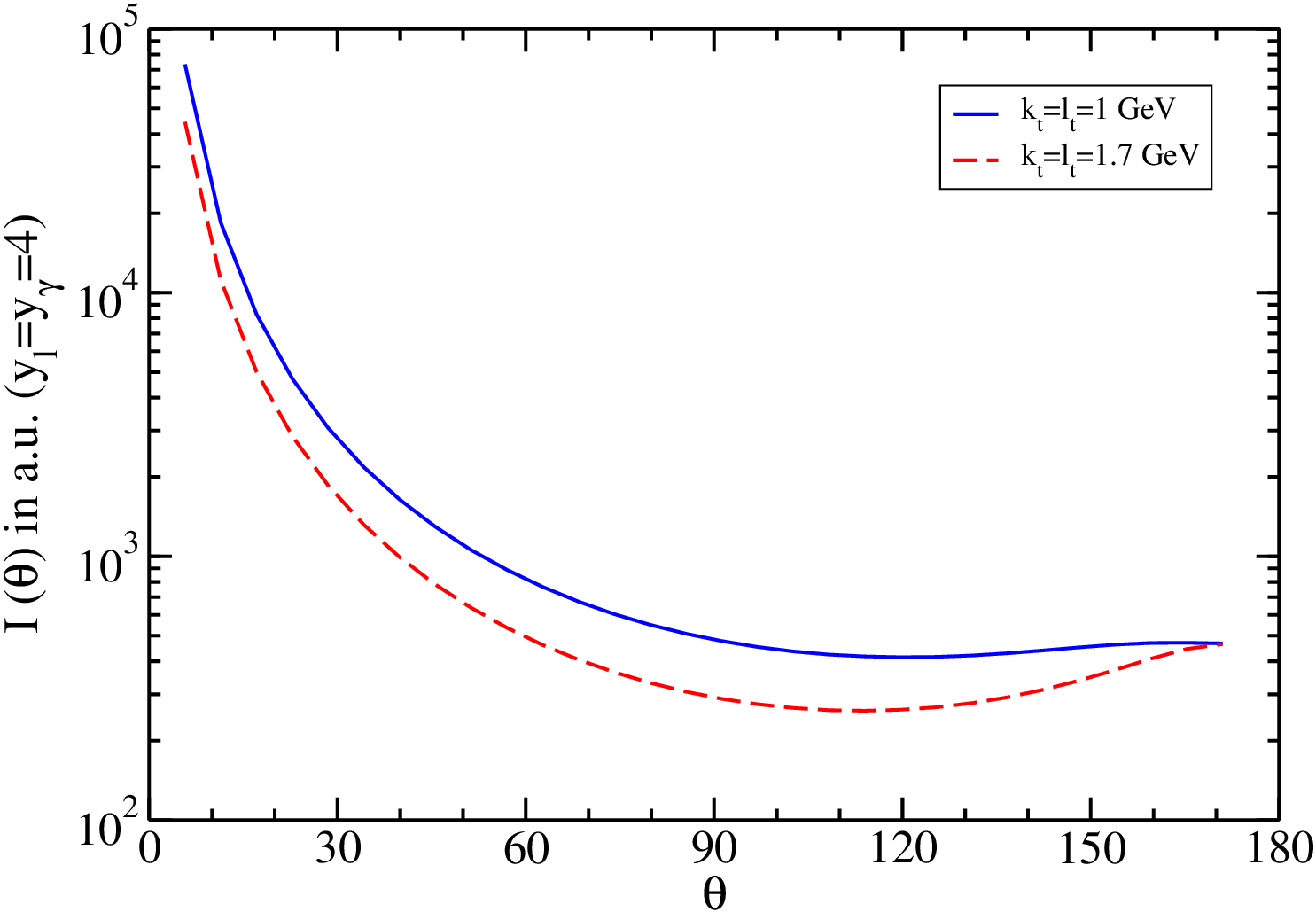}}
\caption{The angular correlation $I (\theta)$  in the forward rapidity 
region $y_l = y_{\gamma} = 4$.}
\label{fig:I_theta_y4}
\end{figure}

To relate our results for the parton production to the case hadron $+$ photon 
production, we need to convolute the partonic cross section in (\ref{eq:cs_gen}) with the 
parton distribution function in a deuteron and the quark-hadron fragmentation
function. The hadron $+$photon production cross section can then be written as
\be
{d\sigma^{d\, A \rightarrow h(q)\,\gamma(k)\, X}
\over d^2b_t\, dk_t^2\, dq_t^2\, dy_{\gamma}\, dy_h\, d\theta} =
\int_{z_{min}}^1 {dz \over z^2} \, \int dx \, f_{q/d} (x, Q^2) \, D_{h/q} (z,Q^2)  
{d\sigma^{q(p)\, A \rightarrow q(l)\,\gamma(k)\, X}
\over d^2b_t\, dk_t^2\, dl_t^2\, dy_{\gamma}\, dy_l\, d\theta} 
\label{eq:cs_had}
\ee
where a sum over different quark and anti-quark flavors is understood
and $y_h, q_t$ are the rapidity and transverse momentum of the produced
hadron. We have neglected hadron masses so that the rapidity of the produced
quark ans hadron are the same. The momentum fraction $z$ is defined as
$z= q_t/l_t = q^-/l^-$ and its minimum value in (\ref{eq:cs_had}) is given by 
$z_{min} \equiv {q_t \over \sqrt{s}} [{ 1 \over 1- {k_t \over \sqrt{s}}}]\, e^{y_h - y_{\gamma}}$.
Since the nuclear modification (shadowing) of the deuteron wave 
function is expected to be small in this kinematics, it is ignored. We 
use the GRV98 set of parton distribution functions \cite{grv98} as well the KKP 
hadron fragmentation functions \cite{kkp}. The integral over $x$ is simple and can be
done using the delta function present in the partonic cross section so that
there is effectively only one integration to perform.

\begin{figure}[htp]
\centering
\setlength{\epsfxsize=8cm}
\centerline{\epsffile{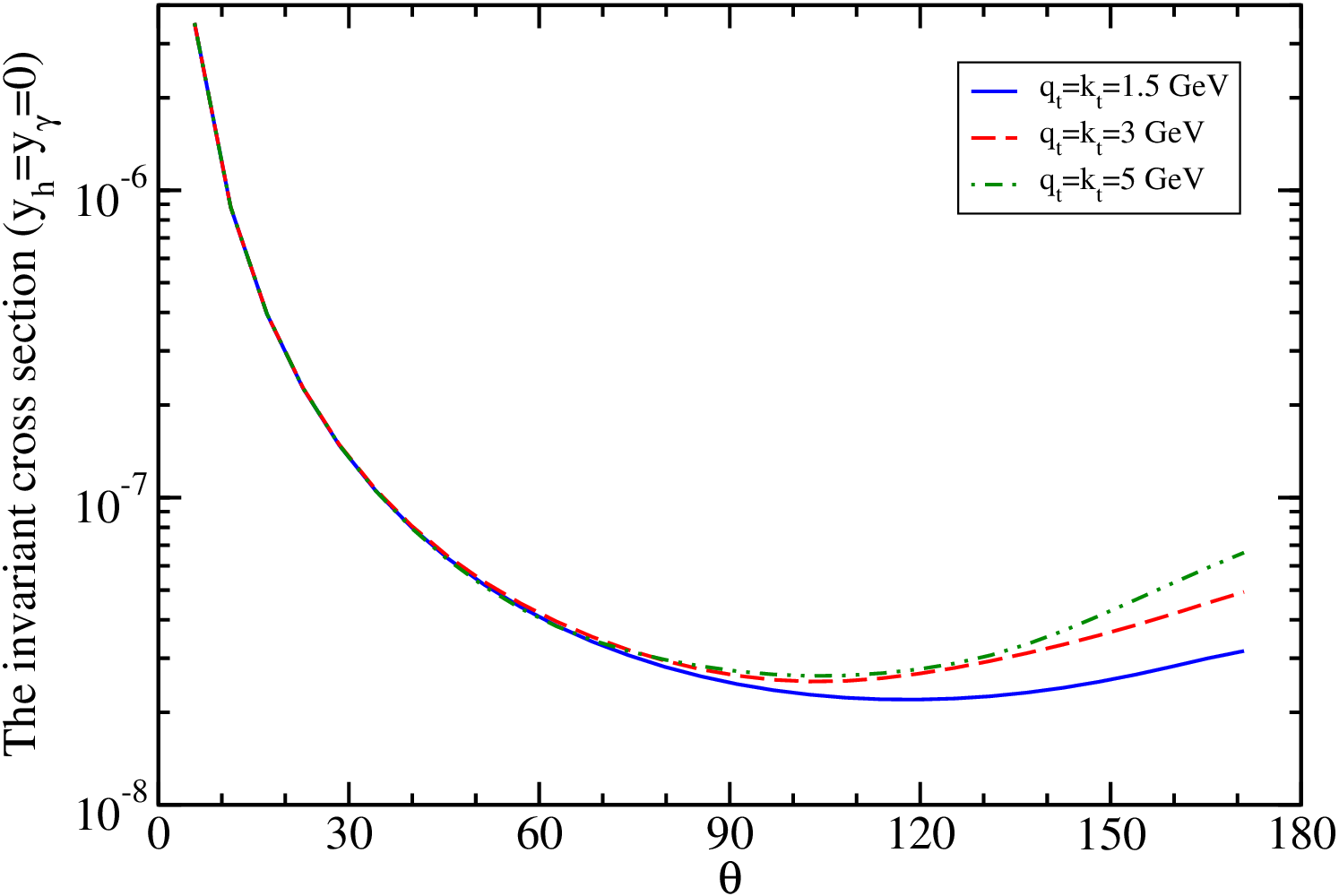}}
\caption{The invariant cross section given by (\ref{eq:cs_had}) at different transverse momenta.}
\label{fig:cs_d_y0}
\end{figure}

In Fig. (\ref{fig:cs_d_y0}) we show the invariant cross section given by (\ref{eq:cs_had}) at
mid rapidity RHIC for three different transverse momenta; 
$q_t=kT= 1.5 GeV,\, q_t=k_t=3 GeV $ and $q_t=k_t=5 GeV$. The lowest 
transverse momentum of the hadron considered is $\sim 1.5 GeV$ since the fragmentation functions 
of KKP start at $Q_0=1.5 GeV$. Again, all the curves are 
normalized to match at the lowest angle for ease of comparison. As is seen, the away side correlation 
is much less for smaller momenta, which is this case is comparable to the saturation scale
of the nucleus $Q_s \sim 1.5 GeV$. The largest considered transverse momentum, $5 GeV$,
is much larger than the saturation scale and most likely, even larger than the so called
extended scaling scale. The cross section is again much larger for the
collinear configuration of the hadron and photon. This will be modified by the isolation
cuts imposed by the future detector setup at RHIC. However, the rise of the correlation with
higher transverse momenta is clear even without any isolation cuts. Furthermore, the isolation
cuts imposed on collinear photons will not affect the large angle correlation even though
it will be much easier to judge whether the away side peak is as large as, or comparable to,
 the near side peak which is not possible to tell without an isolation cut.
\vspace{0.3in}

In Fig. (\ref{fig:cs_d_qtkt}) we show the invariant cross section as given by (\ref{eq:cs_had})
at different rapidities:
\vspace{0.3in}
\begin{figure}[htp]
\centering
\setlength{\epsfxsize=8cm}
\centerline{\epsffile{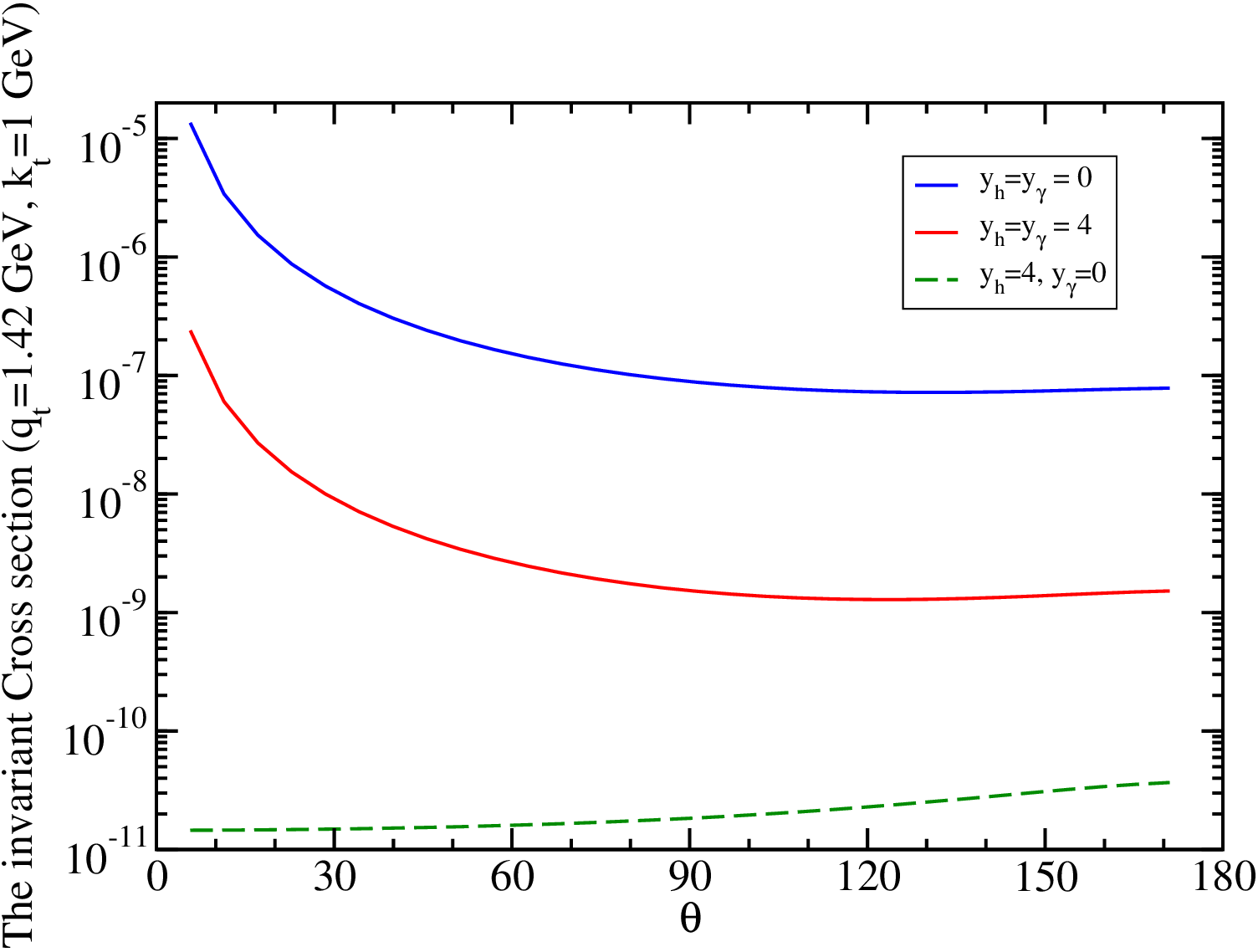}}
\caption{The invariant cross section (\ref{eq:cs_had}) at different rapidities.}
\label{fig:cs_d_qtkt}
\end{figure}
 We consider the cases when both hadron and photon are at the same
rapidity; either mid rapidity ($y=0$) or forward rapidity ($y=4$), and the case when they
are far apart in mid rapidity, i.e. when the photon is in mid rapidity and the hadron is in
the forward region. 
We note that in the kinematics when the two particles have the same transverse 
momenta but are widely separated in rapidity, there is no collinear divergence due to the large
factor of $e^{y_{\gamma} - y_h}$ in the denominator. Physically, this is easy to understood
since two particles widely separated in rapidity can not be collinear for comparable transverse
momenta. When the two produced particles are at the same rapidity, there is always a 
collinear divergence for comparable transverse momenta. We finally note that at rapidity
of $y=4$, we are very close to the edge of the kinematic limit for transverse momenta $\le 2 GeV$
which limits our ability to study the change in the away side peak with increasing transverse 
momenta. Having a more refined hadron or photon detection capabilities in the intermediate 
rapidity region will greatly help with this.

\section{Discussion}
We have calculated the hadron $+$ photon production cross section in high energy
deuteron-nucleus collisions at RHIC energy and investigated the dependence of the
cross section on the angle between the produced hadron and photon. It is shown that
the magnitude of "away" side peak depends sensitively on the transverse momenta
of the particles produced and can thus be a sensitive measure of the importance of gluon
saturation physics. Due to the collinear divergence present in the cross section, one
needs to introduce an isolation cut for the photon. Since this is largely a matter of
detector kinematics and capabilities which will improve with the possible upgrades
at RHIC in the future, we do not include isolation cuts and leave this for future studies.

We used the dipole model of KKT for our quantitative studies. It is probably worthwhile
to do this analysis using the other parameterizations available, once the experimental
setup becomes more clear. The KKT dipole profile has been used successfully
to fir the forward rapidity data without any model assumptions \cite{aaj}. However,
in mid rapidity RHIC, KKT dipole profile seem to require a much steeper anomalous
dimension \cite{aaj2} or extra model assumptions \cite{kkt2}. Therefore, one should be cautious
using the KKT parameterization in mid rapidity RHIC. 

 Furthermore, one can avoid the difficulties of the photon
isolation cut procedure, by using hadron $+$ dilepton production cross section 
in order to study the angular correlations and the role of saturation dynamics. The dilepton
pair invariant mass will regulate the divergence of the photon $+$ hadron cross section
and therefore is a cleaner process to measure even though the rates are smaller.
We leave this for future studies.

Finally, with the expected proton-lead run at LHC coming up in the near future, one will have a 
much broader rapidity coverage which will be helpful in many ways. Due to the larger
center of mass energy compared to RHIC, the saturation scale will be larger at comparable
rapidities so that the saturation effects will be stronger, and therefore, more robust. 
Furthermore, some of the proposed detectors such as CMS are expected to have very
forward measurement capabilities. For example, it may be possible to study correlations at
LHC between particles separated by 5-7 units of rapidity, and at the same time, have
quite a large phase space in transverse momentum available. This will greatly help
our ability to study the different kinematics regions of the Color Glass Condensate by
dialing the rapidity and transverse momenta of the produced particles.

\vspace{0.2in}
\leftline{\bf Acknowledgments} 

The author thanks F. Gelis and W. Vogelsang for helpful discussions and 
F. Gelis for the use of his Fortran code for evaluating the  dipole cross section.
This work is supported in part by the U.S. Department of Energy under Grant No. 
DE-FG02-00ER41132.
     
\vspace{0.2in}
\leftline{\bf References}

\renewenvironment{thebibliography}[1]
        {\begin{list}{[$\,$\arabic{enumi}$\,$]}  % {\arabic{enumi}.}
        {\usecounter{enumi}\setlength{\parsep}{0pt}
         \setlength{\itemsep}{0pt}  \renewcommand{\baselinestretch}{1.2}
         \settowidth
        {\labelwidth}{#1 ~ ~}\sloppy}}{\end{list}}

\end{document}